\documentclass[namedreferences]{solarphysics}
\usepackage[optionalrh,solaenum]{spr-sola-addons} 
\usepackage{graphicx}                    
\usepackage{color}                       
\usepackage{url}                         
\usepackage{soul}                         
\usepackage{epstopdf}

\newcommand{\aap}{    {\it Astron. Astrophys.}}

\newcommand{\apj}{    {\it Astrophys. J.}}

\newcommand{\grl}{    {\it Geophys. Res. Lett.}}

\newcommand{\jastp}{  {\it J. Atmos. Solar Terr. Phys.}}
\newcommand{\jgr}{    {\it J. Geophys. Res.}}

\newcommand{\nat}{    {\it Nature}}

\newcommand{\solphys}{{\it Solar Phys.}}

\newcommand{\ssr}{    {\it Space Sci. Rev.}}

\begin{document}

\begin{article}

\begin{opening}

\title{Detection of Solar Rotational Variability in the LYRA 190\,--\,222 nm Spectral Band}

%
\author{A.V.~\surname{Shapiro}$^{1,2}$\sep
	     A.I.~\surname{Shapiro}$^{1}$\sep
              M.~\surname{Dominique}$^{3}$\sep
              I.E.~ \surname{Dammasch}$^{3}$\sep 
              C.~\surname{Wehrli}$^{1}$\sep
                 E.~\surname{Rozanov}$^{1,2}$\sep
               W.~\surname{Schmutz}$^{1}$
                     }

\runningauthor{\it A.V. Shapiro et al.}
\runningtitle{Detection of Solar Rotational Variability in the LYRA 190\,--\,222 nm Spectral Band}

%
  \institute{$^{1}$ Physikalisch-Meteorologishes Observatorium Davos, World Radiation Center, 7260 Davos Dorf, Switzerland\\
      email: \url{anna.shapiro@pmodwrc.ch}  \\           email: \url{alexander.shapiro@pmodwrc.ch}\\  email: \url{christoph.wehrli@pmodwrc.ch} \\ email: \url{werner.schmutz@pmodwrc.ch} \\
          $^{2}$ Institute for Atmospheric and Climate Science ETH, Zurich, Switzerland \\
            $^{3}$ Royal Observatory of Belgium, Ringlaan 3 B-1180 Brussel, Belgium\\
             email: \url{Marie.Dominique@oma.be} \\ email:\url{dammasch@oma.be} \\ 
             }

\begin{abstract} 
We analyze the variability of the spectral solar irradiance during the period from 7 January, 2010 until 20 January, 2010 as measured by the Herzberg channel (190\,--\,222 nm) of the {\it Large Yield RAdiometer} (LYRA) onboard PROBA2. In this period of time observations by the LYRA nominal unit experienced degradation and the signal produced by the Herzberg channel frequently jumped from one level to another. Both these factors significantly complicates the analysis. We present the algorithm which allowed us to extract the solar variability from the LYRA data and compare the results with SORCE/SOLSTICE measurements and with modeling based on the {\it Code for the Solar Irradiance} (COSI).
\end{abstract}

%

\end{opening}

\section{Introduction.}\label{sect:intro} 
After the start of regular space-born measurements in 1978, it became clear that the solar irradiance is variable on different time-scales. Since then significant progress was achived both in measuring and in modeling of the solar irradiance variability \cite{froehlich2005,krivovasolanki2008}. 
At the same time, new missions devoted to monitoring of the spectral solar irradiance (SSI) continue to bring significant surprises  (\opencite{harderetal2009}). So our  understanding of the  mechanisms of the solar irradiance variability depends on observations by new instruments. The importance of studying the solar irradiance variability is further emphasized by its direct impact on the Earth's climate \cite{haighetal2010,grayetal2010}.

In this article we analyze  the  spectral solar irradiance data from the recent European mission PROBA2 (launched on 2 November, 2009). 
LYRA is a solar radiometer onboard PROBA2, which is a technologically oriented ESA micro-mission, and is observing the solar irradiance in two UV and two EUV spectral channels  \cite{LYRA1,LYRA2,marie2011}. The passbands of the UV channels  were selected for their relevance for the  ozone concentration. 

Up to now the LYRA data were used only for the analysis of variations shorter than one day.  \inlinecite{flare_LYRA} studied flares observed by LYRA. \inlinecite{occultations_LYRA} and \inlinecite{eclipse_paper} analyzed the light curves during occultations and eclipses, respectively. In this article we analyze the variations of the solar irradiance during a period of approximately two weeks, which corresponds to the transit of active regions across the solar disk and is representative for studying the solar rotational cycle \cite{fliggeetal2000}. 
We compare our  results with SORCE/SOLSTICE ({\it SOLar-STellar Irradiance Comparison Experiment}, \opencite{SOLSTICE_design}). We also use COSI (\opencite{shapiroetal2010}) as a tool for modeling the variability of the irradiance, assuming that the latter is determined by the evolution of the solar surface magnetic field. The theoretical results are compared with the LYRA measurements. Our analysis is restricted to the Herzberg channel of LYRA as  irradiance in the Herzberg continuum (190\,--\,222 nm) presents challenge for the solar radiative transfer models \cite{eclipse_paper} and is especially important for climate modeling \cite{brasseuretal1997,rozanovetal2006,anna2011}. 

The analysis of the solar variability in this channel is significantly hampered by the severe degradation, which LYRA started to experience immediately after the covers had ben opened. The degradation led to a significant loss of sensitivity to the solar signal already after a month since first light (6 January, 2010). This is why, for the current analysis, we used early data though the instrument was still in its commissioning phase. At that time, the spacecraft was slightly off-pointing, due to a problem in its onboard software which was fixed at the end of January 2010. Another significant problem was therefore pointing fluctuations which in combination with inhomegenous sensitivity of the diamond detectors resulted in significant fluctuations of the data. We note that the amplitude of the signal introduced by the jitter of the PROBA2 exceeds the natural solar variability. The data were also corrupted by the angular rotation of PROBA2 (four times per orbit), occultations, electronic perturbations, and sudden jumps (which are specific for the Herzberg channel and are not yet fully understood). The latter problems could usually be easily identified and corrected, so they did not present such important difficulties as degradation and pointing fluctuations. The detailed discussion of the different perturbations in the LYRA data are given by \inlinecite{marie2011}.

We analyze the LYRA data for the period from 7 January, 2010 until 20 January, 2010. During this period one sunspot group surrounded by a plage region made a full transit across the near-side of the Sun (see Figure~\ref{fig:Sun}).   At the same time there were no other significant active regions on the solar disk, so the variability pattern was expected to have a relatively simple profile. In Section ~\ref{sect:LYRA} we present the analysis of the LYRA data and show that although the accuracy of the processed signal is low, we believe that we have extracted the rotational signature of the Sun. The signal is compared with SORCE/SOLSTICE measurements and COSI calculations in Section ~\ref{sect:COSI}. We summarize the result in Section~\ref{sect:conc}.

\section{Analysis of the LYRA data}\label{sect:LYRA}
\begin{figure} 
\centerline{\includegraphics[width=1.0\textwidth,clip=]{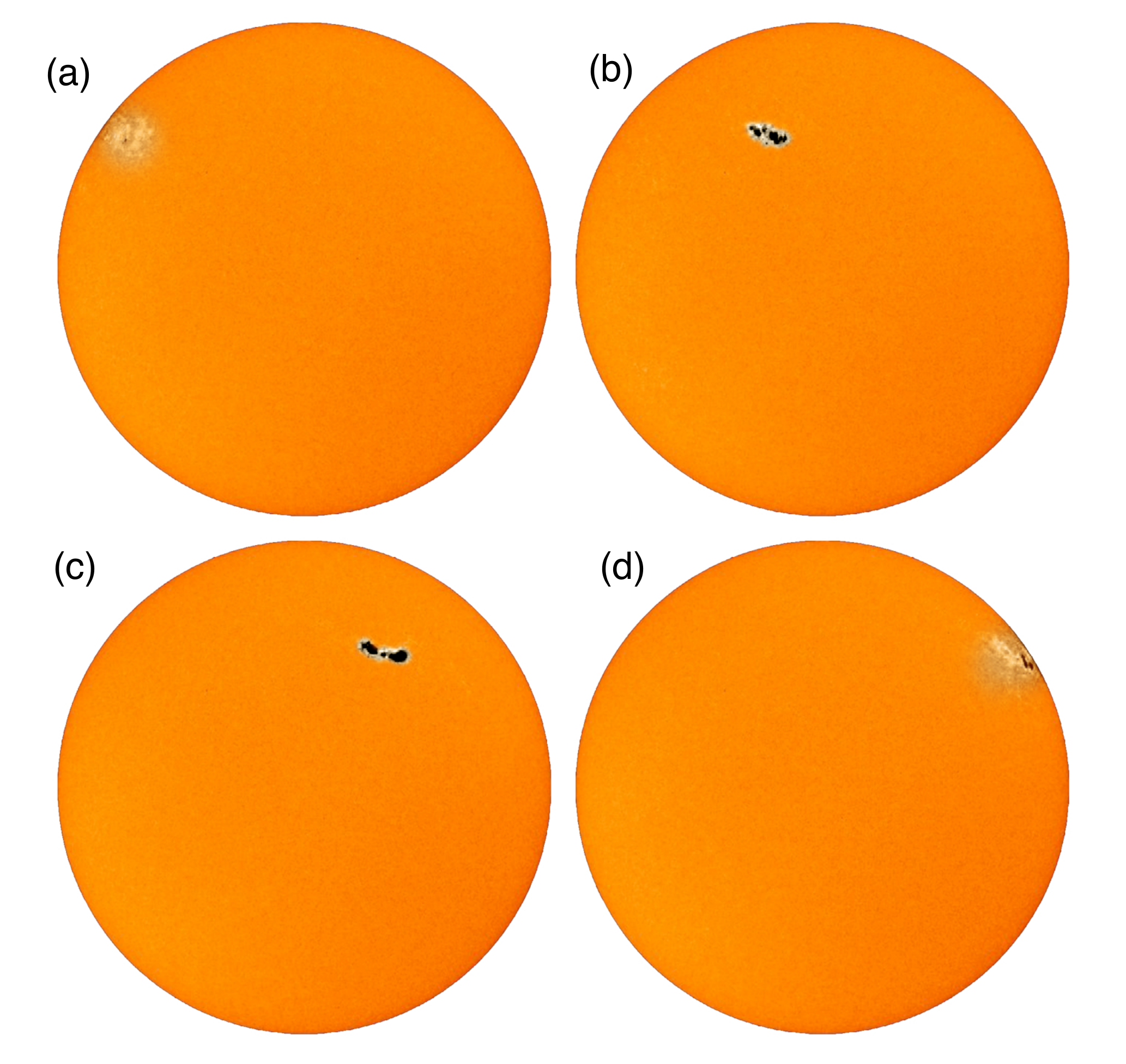}}
\caption{SOHO/MDI continuum images for 8 January (a), 11 January (b), 14 January (c), 17 January (d) of 2010. For better clarity the contrasts between sunspot, plage, and quiet Sun were artificially increased.}
\label{fig:Sun}
\end{figure}
The sunspot group surrounded by plage appeared on the solar disk on 7 January, 2010 and disappeared on 20 January, 2010. The transit is shown on Figure~\ref{fig:Sun}. One can expect that the presence of active regions on the solar disk will modify the solar irradiance \cite{fliggeetal2000}. In this section we present the method that was used to extract these modulations from the LYRA data.

\begin{figure} 
\centerline{\includegraphics[width=1.0\textwidth,clip=]{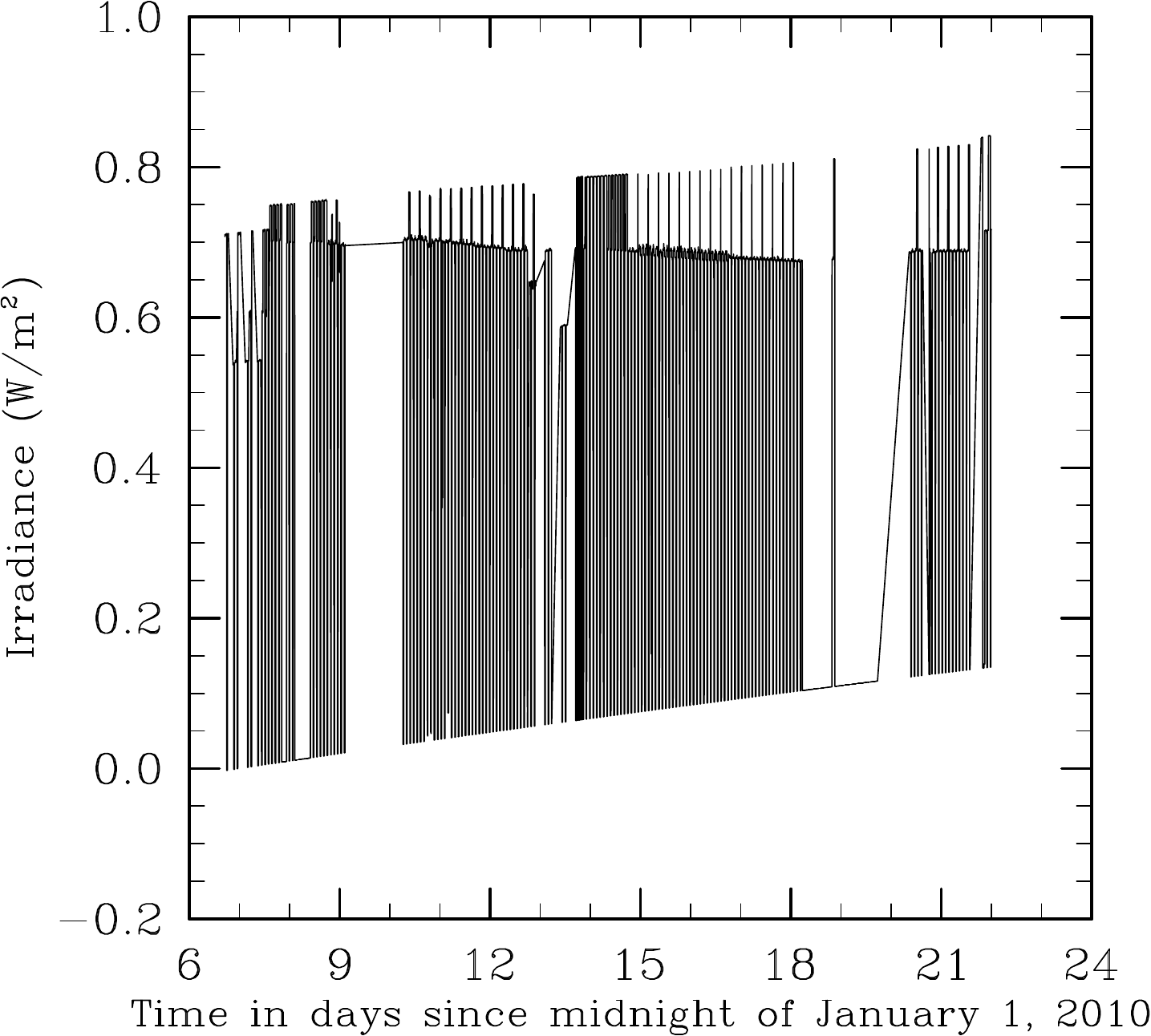}}
\caption{Irradiance measured by the Herzberg channel of LYRA for the 6 January -- 24 January, 2010 period. Plotted are the level3 calibrated data. }
\label{fig:data1}
\end{figure}

The level3 calibrated data from the Herzberg channel of LYRA for January 2010 are plotted in Figure~\ref{fig:data1}. These data are available for the community  \cite{marie2011}  and corrected for the temperature effects, degradation, dark current, and one minute averaged. The LYRA samples in Figure~\ref{fig:data1} can be sorted in two groups: the bottom line corresponds to data acquired in occultations by the Earth (dark current), the upper series constitute the actual Herzberg timeseries, but is perturbed by jumps induced by the electronics (appearing when the FPGA was reloaded) and by pointing fluctuations. The selection of valid data was therefore not trivial, and we had to process them with a special care.

\begin{figure} 
\centerline{\includegraphics[width=1.0\textwidth,clip=]{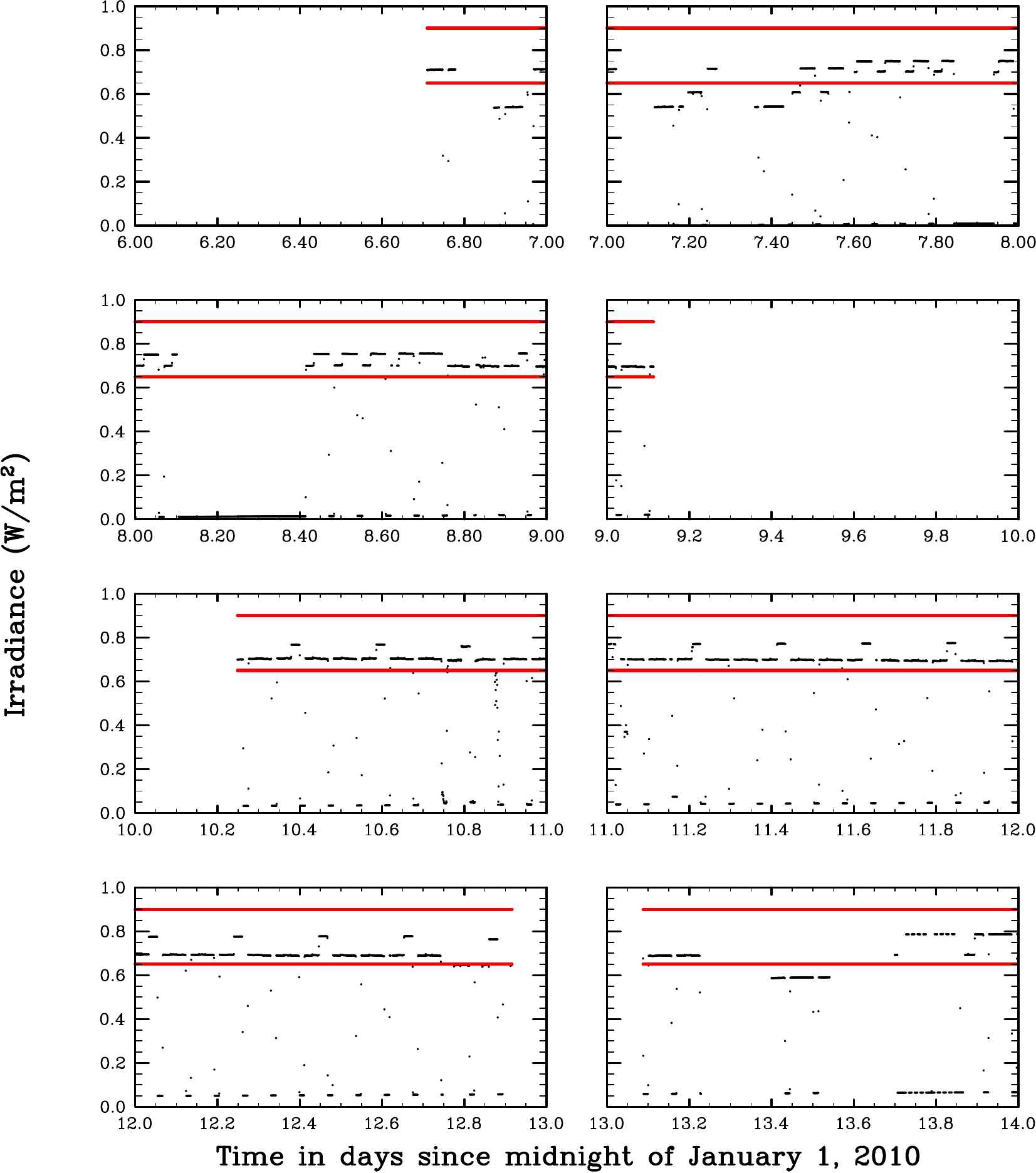}}
\caption{Irradiance measured by the Herzberg channel of LYRA for the 6 January -- 14 January, 2010 period. The red solid lines denotes the trustable range of the irradiance which was used for the analysis.}
\label{fig:data2}
\end{figure}

\begin{figure} 
\centerline{\includegraphics[width=1.0\textwidth,clip=]{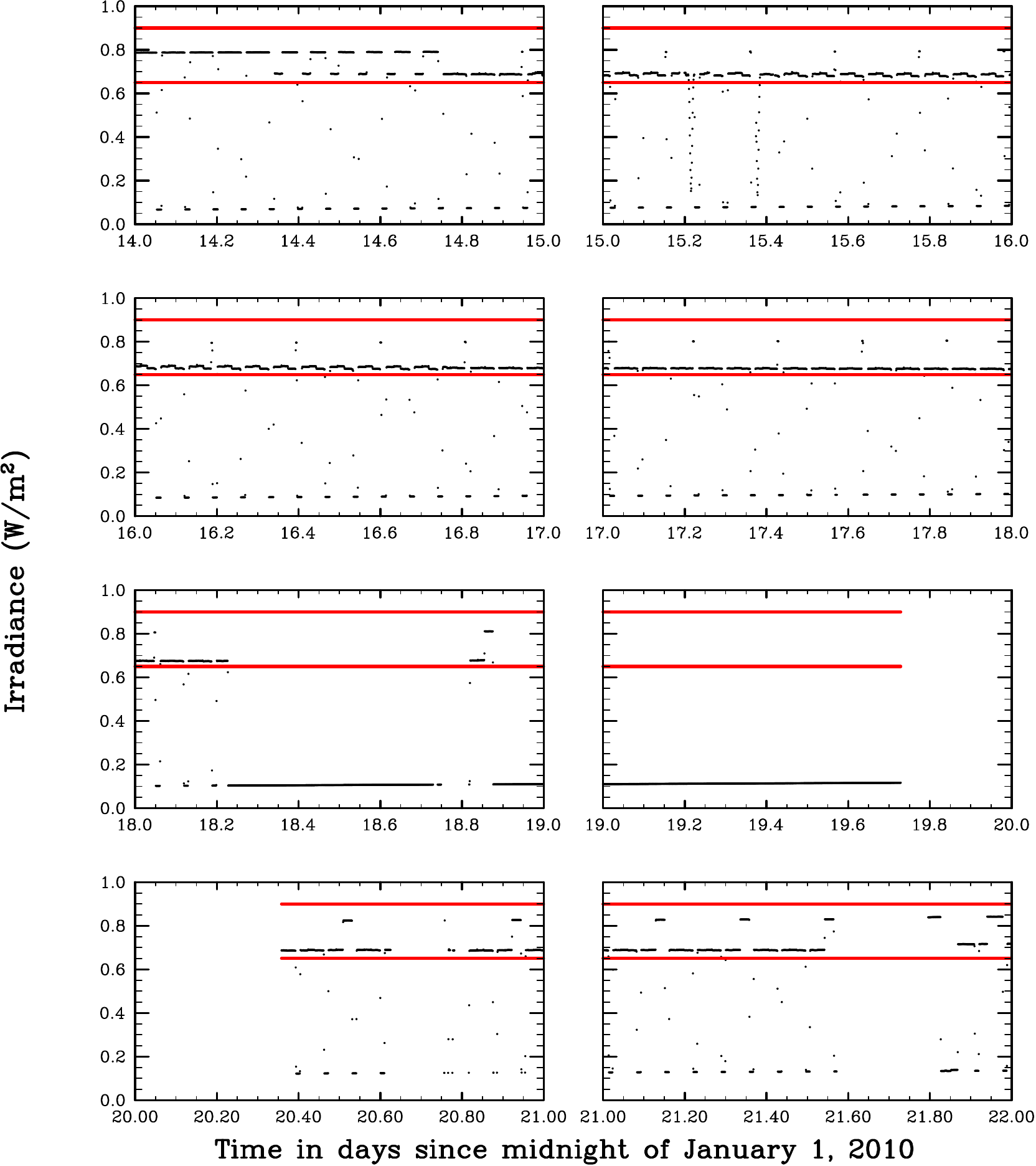}}
\caption{Irradiance measured by the Herzberg channel of LYRA for the 14 January -- 22 January, 2010 period. The red solid lines denotes the trustable range of the irradiance which was used for the analysis.}
\label{fig:data3}
\end{figure}

\begin{figure} 
\centerline{\includegraphics[width=1.0\textwidth,clip=]{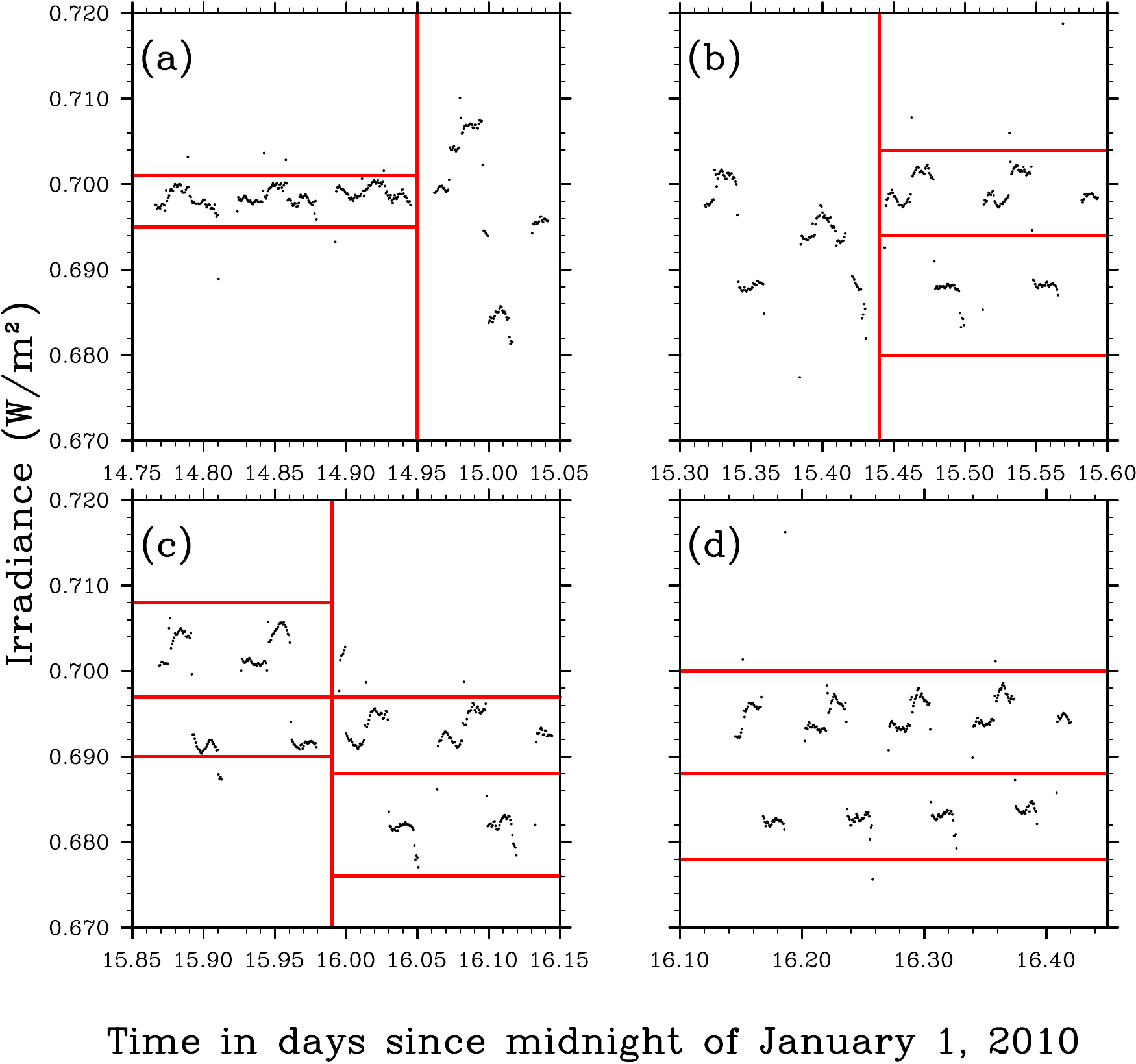}}
\caption{Level3 calibrated LYRA data in the Herzberg channel with high temporal resolution. The red solid lines restrict different ``branches'' of the irradiance level.}
\label{fig:data4}
\end{figure}

\begin{figure} 
\centerline{\includegraphics[width=1.0\textwidth,clip=]{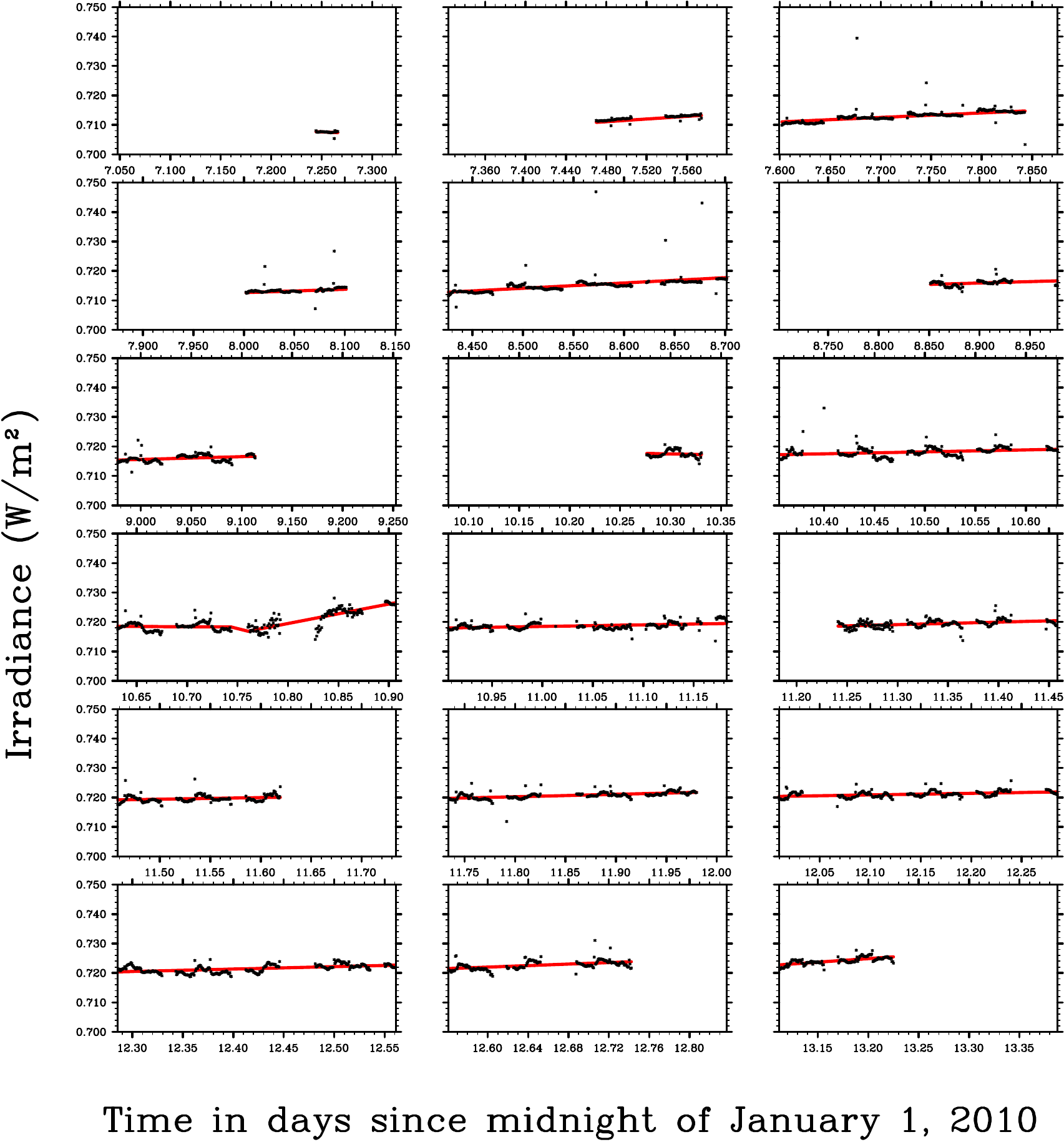}}
\caption{The implementation of the linear regression (red lines) through the four-orbit intervals of the LYRA data for the 7 January -- 13 January, 2010 period.}
\label{fig:data5}
\end{figure}

\begin{figure} 
\centerline{\includegraphics[width=1.0\textwidth,clip=]{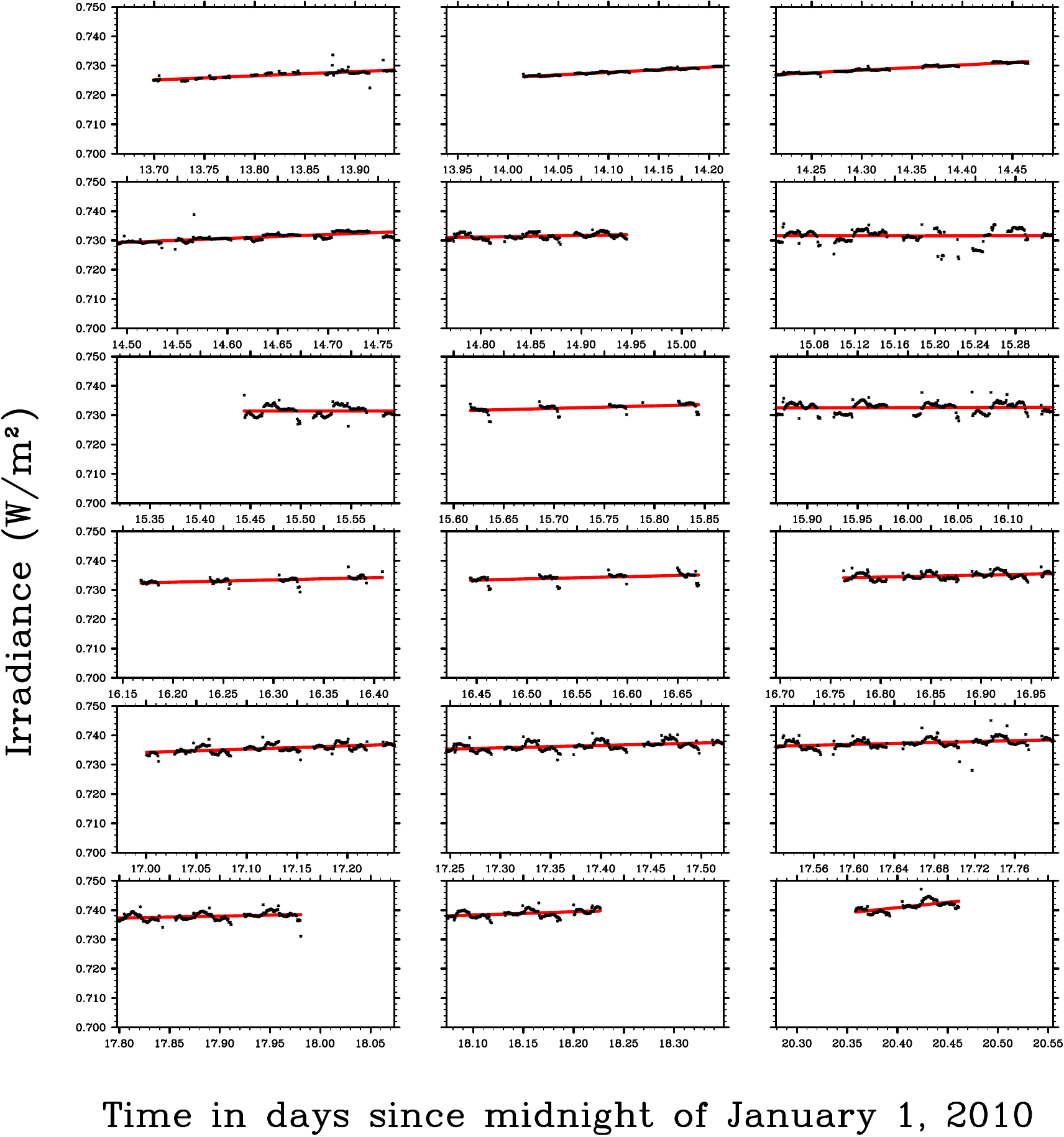}}
\caption{The implementation of the linear regression (red lines) through the four-orbit intervals of the LYRA data for the 13 January -- 20 January, 2010 period.}
\label{fig:data6}
\end{figure}

 The part of the data corresponding to periods of significant pointing fluctuations or occultations had to be excluded from the analysis. Therefore in the first step we chose a trustable range of the irradiance and excluded the outliers.  For the period under consideration (January 2010 ) the lower and upper  levels were chosen to be 0.65 Wm$^-$$^2$ and  0.9 Wm$^-$$^2$, respectively. This procedure is illustrated in Figures~\ref{fig:data2} and \ref{fig:data3}. One can see that even within the selected trustable range the irradiance level is highly unstable and undergoes a few jumps per day. For some of the days several distinctive ``branches''  can be clearly seen. The irradiance level is stable within each of the branches but constantly  jumps between them.  Such behavior can be attributed to pointing fluctuations and rotation of PROBA2, as well as to some electronic perturbations.  

The upward slope in the zero level of the irradiance is due to a constant which was added to remove the degradation for  the production of the level3 data.  The additive correction of the degradation is justified for the analysis of the flares but is not suitable for the analysis of the rotational cycle, where a multiplicative correction is preferable (see \opencite{marie2011}, \opencite{eclipse_paper}). Therefore we had to subtract this constant from the level3 data. To correct for the degradation, we calculated the change of the irradiance at the Herzberg channel  between 7 January and 18 January as measured by SORCE/SOLSTICE. The SOLSTICE data were convolved with the profile of the LYRA Herzberg channel (\opencite{LYRA2}). Then, following the approach of \inlinecite{marie2011} for the second half year of LYRA observations, we tested two different corrections for the degradation, one of the type  ${\rm \frac{1}{a+b*time}}$, and another of the type ${\rm \exp(a-b*time)}$. A spline correction was not used as we are interested only in a relatively short period of time.  In contrast to \inlinecite{marie2011} we applied the corrections not as an additive but as a multiplicative factor. In both cases we chose the coefficients $a$ and $b$ so that the change of the irradiance between 7 January and 18 January was the same in LYRA and SOLSTICE data. No substantial difference was seen between the data corrected with linear and exponential functions so we only show the data processed with a linear correction.

The segregation of  the LYRA time series to different branches is further illustrated in Figure~\ref{fig:data4}, which shows four selected time intervals with high  temporal resolution. The data plotted in Figure~\ref{fig:data4}(a) consists only of one branch which covers the 14.75 -- 14.95 time interval. The data in the 14.95 -- 15.05 interval were considered to be too noisy and excluded from the analysis. The data plotted in Figure~\ref{fig:data4}(b) consist of two branches (both in 15.44 -- 15.60  interval), while the data in the 15.30 -- 15.44 interval were excluded. The data plotted in Figure~\ref{fig:data4}(c) consist of four branches (two branches prior to 15.99 and two after), while the data in Figure~\ref{fig:data4}(d) consist of two branches covering the entire time interval.

The structure of the LYRA data significantly complicates the analysis as the amplitude of the jumps is sometimes larger than expected amplitude of the solar variability. The fundamental assumption of our analysis is that, although the  variability of the LYRA data is dominated by these jumps, every individual branch does have a signature of the solar variability. 

To extract this signature, we binned the available data to the four orbit intervals (see Figures~\ref{fig:data5} and \ref{fig:data6}).  Each interval starts when PROBA2 passes the Earth's equatorial plane and lasts $4 \times 100$ = 400 minutes. In each of these intervals we took the first stretches of every branch (16.142 -- 16.162  and 16.162 -- 16.182 intervals on Figure~\ref{fig:data4}(d)) and calculated the mean values of the irradiance at these stretches. Then we shifted one (or more if the interval contained more than two branches)  of the branches enforcing the condition that these mean values should be equal to each other. The amplitude of the jumps did not change significantly within the four-orbit interval (which also confirms that jumps occur due to the fact that PROBA2 needs to rotate four times per orbit), so this  allowed us to eliminate all of the jumps from the data. The procedure described above is basically equivalent to the correction of the flat field. 
In principle one can expect that the mean values of the irradiance at the considered stretches could be different due to the solar variability. However the duration of almost all stretches was very small (Figure~\ref{fig:data4}) so neglecting the variations of the solar irradiance between them does not lead to a significant error.

After the jumps in the data were corrected we applied a linear regression to all available data for each of the four-orbit intervals  (the red lines in the Figures~\ref{fig:data5} and \ref{fig:data6}). The slope of the regression was used to calculate the change of the irradiance level during each of the intervals. In many cases the data did not cover the entire interval and contained a significant gaps in the beginning or in end of the interval. For simplicity we assumed that the level of the solar irradiance did not change during these gaps, so the change of the irradiance corresponds to the projection of the red lines in Figures~\ref{fig:data5} and \ref{fig:data6} to the vertical axis.  An alternative would be an extrapolation of the linear trend, however such extrapolation can significantly increase errors caused by the noise in the data. 

Our final product is a single value of the irradiance per four-orbit interval. The differences between each of the consecutive values were calculated employing the algorithm described above. Let us note that  the implementation of this algorithm violates the condition of the equal change of the irradiance between 7 January and 18 January as measured by  LYRA and SOLSTICE and thus the correction for the degradation should be slightly readjusted. This readjustment  was not yet implemented in Figures~\ref{fig:data5} and \ref{fig:data6} so almost all data in these figures show the upward trend.

The data in 15.04 -- 15.32 and 15.87 -- 16.15 time intervals were too noisy even after the corrections. At the same time the irradiance significantly decreases during these intervals, which can therefore be just an artifact caused by noise. To take this into account we produced two different datasets of the LYRA data.  The LYRA version1 dataset was produced taking all data into account, while LYRA version2 dataset was produced excluding the two aforementioned intervals from the analysis and assuming that the solar irradiance did not change during these intervals. The difference between these datasets indicates the accuracy of our analysis.

\section{Modeling with COSI}\label{sect:COSI}
\begin{figure} 
\centerline{\includegraphics[width=1.0\textwidth,clip=]{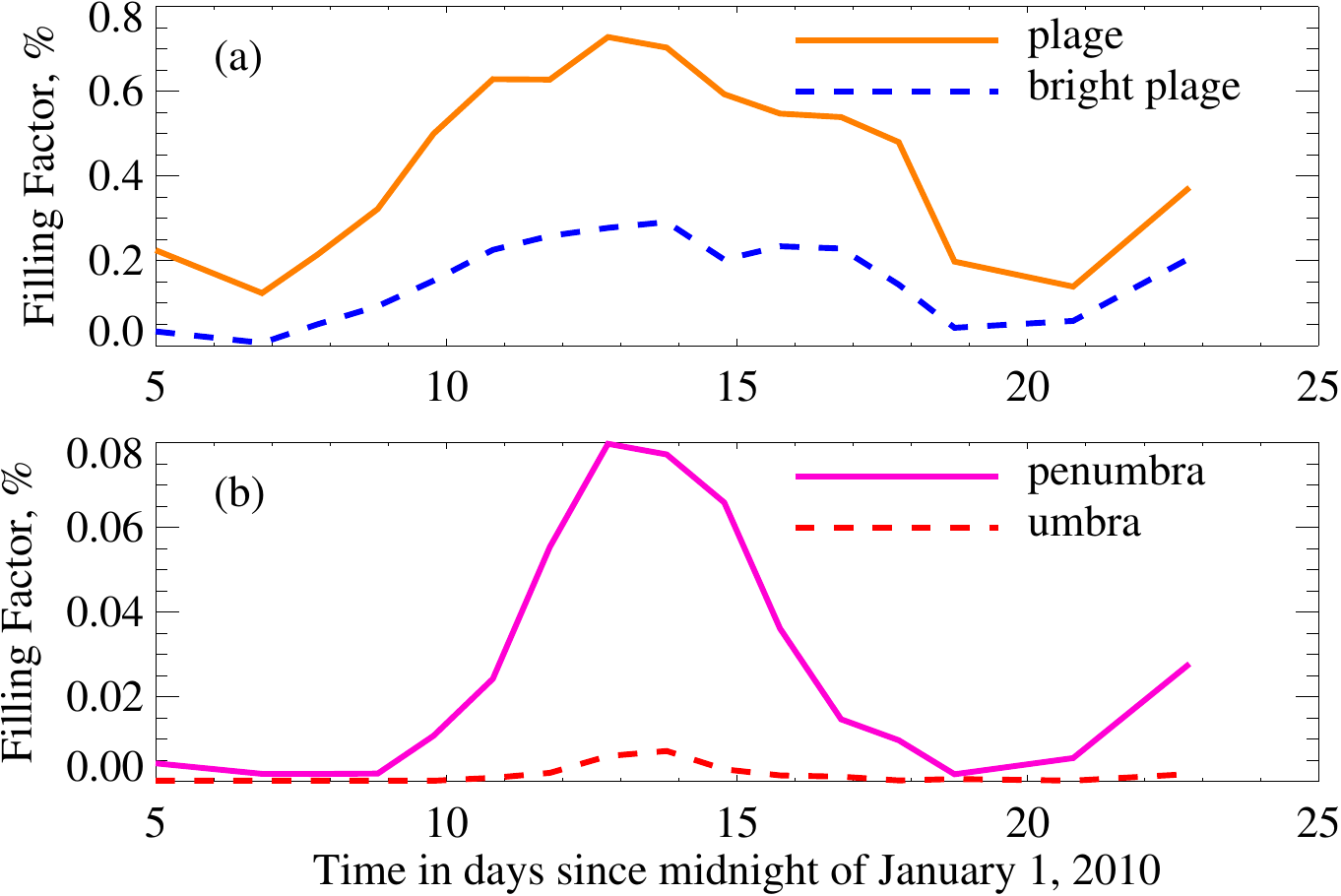}}
\caption{Filling factors of the plage and  bright plage (a), and penumbra and umbra (b) as a function of time. The filling factors are extracted from PSPT images.}
\label{fig:FF}
\end{figure}

\begin{figure} 
\centerline{\includegraphics[width=1.0\textwidth,clip=]{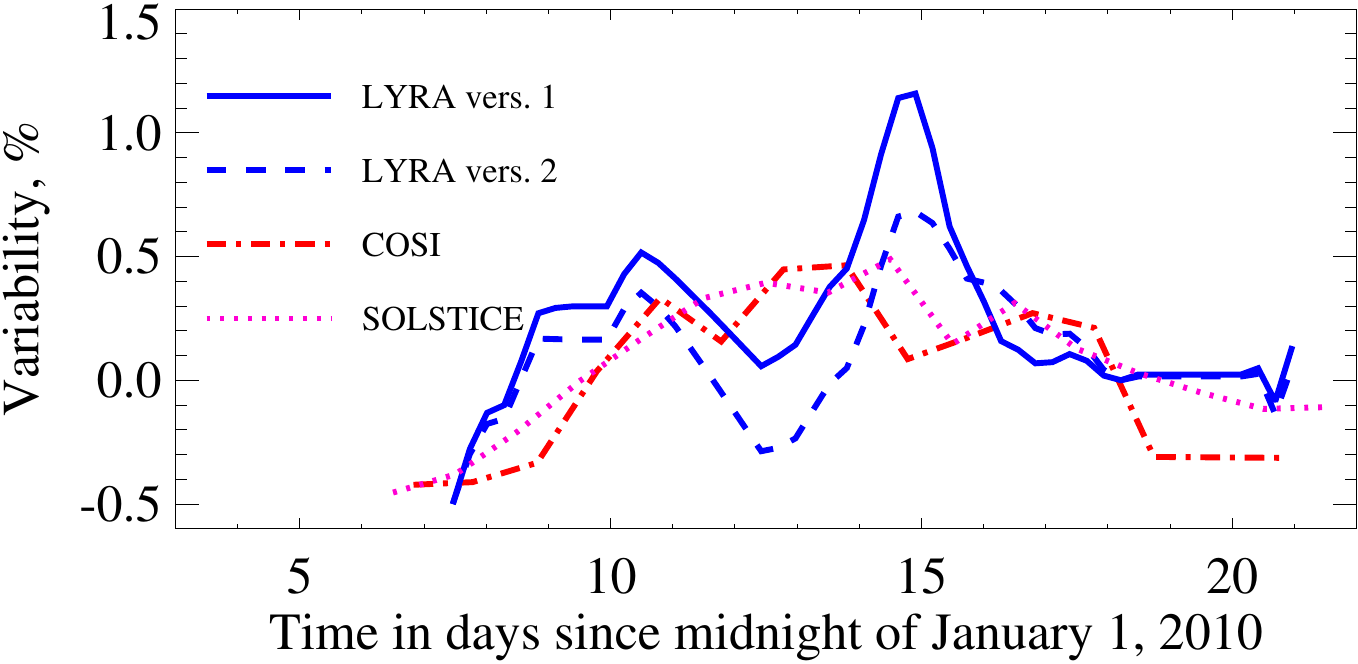}}
\caption{{The variability of the solar irradiance at the Herzberg continuum range (190\,--\,222 nm) as measured by LYRA (versions 1 and 2, see Section~\ref{sect:LYRA}) and SOLSTICE and modeled with COSI.} }
\label{fig:irrad_LYRA}
\end{figure}
In this section we calculate the synthetic profile of the spectral solar irradiance variability for the period analyzed in Section~\ref{sect:LYRA}. We follow a well-developed approach (\opencite{foukallean1988}, \opencite{fliggeetal2000}, \opencite{krivovasolanki2008}, \opencite{domingo2009}, \opencite{SSIrec}) and calculate the time-dependent solar spectrum  as a sum of the spectra from the quiet Sun  and different active features. We employ a four-component model which treats separately contributions from the quiet Sun, sunspots, active network, and plage areas. According to this model,  the solar spectrum $I(\lambda,t)$ can be written as
\begin{eqnarray}
 I(\lambda,t) &=&    \sum\limits_k  \left ( \alpha_{\rm QS} (\mu_k,t) I_{\rm QS} (\lambda,\mu_k) + \alpha_{\rm S} (\mu_k,t) I_{\rm S} (\lambda,\mu_k)    +   \right. \nonumber \\
   &+&  \left. \alpha_{\rm AN} (\mu_k,t) I_{\rm AN} (\lambda,\mu_k)    +  \alpha_{\rm P} (\mu_k,t) I_{\rm P} (\lambda,\mu_k)  \right ) ,
\end{eqnarray}
where $\alpha_{\rm QS} (t)$,   $\alpha_{\rm S} (t)$,  $\alpha_{\rm AN} (t)$,  $\alpha_{\rm P} (t)$ are the time-dependent filling factors of the quite Sun, sunspots, active network, and plage areas accordingly. 
$I_{\rm QS} (\lambda)$,  $I_{\rm S} (\lambda)$, $I_{\rm AN} (\lambda)$, and $I_{\rm P} (\lambda)$ are corresponding synthetic spectra. The summation represents  the division of the solar disk in several concentric  rings and is done over the different heliocentric angles, $\mu_k$ is the corresponding cosines. Following the algorithm of \inlinecite{eclipse_paper} we used thirteen rings. This provides the accuracy of the order of hundredth of percent. 

The synthetic spectra $I_{\rm QS} (\lambda,\mu_k)$, $I_{\rm S} (\lambda,\mu_k)$, $I_{\rm AN} (\lambda,\mu_k)$, and $I_{\rm P} (\lambda,\mu_k)$ are calculated with COSI. The temperature and density structure of the corresponding components are taken from \inlinecite{fontenlaetal1999}. A self-consistent simultaneous solution of the radiative transfer and the statistical equilibrium  equation for the level populations guarantees that COSI considers the correct physics for the Herzberg region where assumption of local thermodynamical equilibrium breaks down. The calculations with COSI yield the spectral solar irradiance which agrees well  with the  SOLSPEC measurements during the ATLAS 3 mission \cite{Grec}.

The time-dependent filling factors were extracted from the {\it Solar Radiation Physical Modeling} (SRPM: \opencite{fontenlaetal1999}, \opencite{fontenlaetal2009}) image mask of the {\it Precision Solar Photometric Telescope} (PSPT: \opencite{rastetal2008}). In Figure~\ref{fig:FF} we give the dependency of the total (summed over all $\mu_k$ values) filling factors of the bright plage, plage, umbra, and penumbra on time. The profile of the dependencies represents the appearance and disappearance of the active features on the solar disk due to rotation (Figure~\ref{fig:Sun}) as well as the projection effect.

The comparisons of the LYRA data with SOLSTICE measurements and calculations are presented in the upper panel of Figure~\ref{fig:irrad_LYRA}. The SOLSTICE  data (available with 1\,nm spectral resolution) and the calculated irradiance (available with 5\,m{\AA} spectral resolution)  were converted with the combined profile of the Herzberg filter and detector (\opencite{LYRA2}). 

Our four-component model of the solar variability is based on the models of the different solar-atmosphere components  from \inlinecite{fontenlaetal1999} who do not distinguish between the bright plage and plage as well as between the umbra and penumbra. At the same time the PSPT filling factors are based on the  \inlinecite{fontenlaetal2009}  where these models are treated separately. To take this into account, we decreased the contrasts of the plage and sunspot with the quiet Sun in a way that the calculated variability in the Herzberg channel matches the SOLSTICE observations. 

The theoretical understanding, as modeled by COSI, suggests  that the variability in the Herzberg channel during the considered period has a one-peak profile. This can be explained by the dependency of the contrast between the different components of the solar atmosphere on the wavelength and heliocentric angle. 
The contrast between the bright components of the solar atmosphere (plage and active network) and quiet Sun strongly increases towards shorter wavelengths, while the contrast between the quiet Sun and sunspot depends on the wavelength more gradually (Figures 11 and 12 from \opencite{shapiroetal2010}).  As a result, for the considered transit of the active regions the increase of the Herzberg irradiance due to the presence of the plage and active network overweight the decrease of the Herzberg irradiance due to the presence of the sunspot independent of their  position on the solar disk (for a more detailed explanation see \opencite{unruhetal2008}). The difference between LYRA and SOLSTICE or COSI indicates the uncertainty in the current processing of LYRA data.

\section{Conclusions}\label{sect:conc}
The LYRA data in the Herzberg channel of the nominal unit are very much disturbed by the degradation and inhomogeneous flat field of the detector. However with careful and laborious analysis the real solar signal can be extracted.  We believe that the processing of the data presented in this article is robust. We should also note that the main objective of this article should be seen not in the physical analysis of the solar variability (which is hindered by the aforementioned problems) but rather in the presenting the algorithm for the extracting the signatures of the solar variability from such noisy time series. Even if the LYRA data are not be used in future, this algorithm can to some extent be useful for the analysis of datasets with the similar problems.

The solar signal extracted from the LYRA data is in reasonable agreement with SORCE/SOLSTICE measurements (Figure~\ref{fig:irrad_LYRA}). However the LYRA measurements indicate significant increase of the irradiance on 13 and 14 January, while there is no such increase in the SOLSTICE data. The theoretical results agree better with SOLSTICE measurements  and yield a one-peak profile of the variability for the time period under consideration. 

We believe that the origin of this disagreement can be clarified analyzing the data from the other recent European mission  PICARD launched on 15 June, 2010.
The PREMOS package onboard PICARD comprises two experiments, one observing solar irradiance in five (two UV, one visible, and two near infrared) spectral channels with filter radiometers, the other measuring TSI with absolute radiometers. One of the PREMOS channels also measures the solar irradiance in the Herzberg continuum range (190\,--\,222 nm) but has a different type of detector  \cite{schmutzetal2009}. The intercomparison of the PROBA2/LYRA and PICARD/PREMOS results will be addressed in a forthcoming article.

\begin{acks}
The research leading to this article was supported by the Swiss National Science Foundation under grant CRSI122-130642 (FUPSOL) and grant 200020-130102. We thank the  PROBA2/LYRA science team for their work in producing the data sets used in this article and their helpful recommendations. We also thank the PSPT team for providing the filling factors of the active components of the solar atmosphere. We thank Stephanie Ebert for her help and advise in editing the figures.
\end{acks}

\bibliographystyle{spr-mp-sola}

\end{article} 
\end{document}